# A method for comparing inferred evolutionary accumulation dynamics across covariates and model structures

Jakob Fernø[1], Michelle B. Verstraaten[2], Lila Gravellier[2], Ellen C. Røyrvik[2,3], Ramon Diaz-Uriarte[4,5], Iain G. Johnston[1,6,*]

1. Department of Mathematics, University of Bergen, Bergen, Norway
2. Department of Clinical Science, University of Bergen, Bergen, Norway
3. Norwegian Institute of Public Health, Bergen, Norway
4. Instituto de Investigaciones Biomédicas Sols-Morreale (IIBM), UAM-CSIC, Madrid, Spain
5. Department of Biochemistry, Universidad Autónoma de Madrid (UAM), Madrid, Spain
6. Computational Biology Unit, University of Bergen, Bergen, Norway

In this note we describe a method for comparing inferred dynamics in evolutionary accumulation models (EvAMs). These models involve the acquisition of multiple, potentially codependent, binary features over time – for example, mutations in cancer development, or phenotypes in evolutionary biology. As the set of methods for inferring EvAM dynamics expands, approaches for comparing inference across algorithms, datasets, and covariates are required. In particular, a comparison method supporting reversible, stochastic dynamics, interactions between feature sets, and potentially non-independent samples (as well as simpler cases) has yet to be established. It is possible for EvAM with similar relative feature orderings to produce completely different sets of observed states due to "frameshift"-like differences; methods distinguishing state and transition similarity are therefore also desirable. Here we suggest a method focussed on ordering matrices, describing the probability that a feature is acquired under different conditions on other features, and is thus generally comparable across methods and datasets. We demonstrate how the approach captures both statistical robustness (given EvAM uncertainty) and scientifically meaningful differences in inferred dynamics. The method is applied to synthetic and real-world data on the evolution of chromosomal aberrations in different tumour types and of drug-resistant bacteria in different countries.

## Introduction

Evolutionary accumulation models (EvAMs) describe how systems accumulate collections of binary features over time (Diaz-Uriarte & Herrera-Nieto, 2022; Diaz-Uriarte & Johnston, 2025). Problem domains include cancer progression, where features may be mutations observed in tumours or single cells (Beerenwinkel et al., 2015; Beerenwinkel & Sullivant, 2009); drug resistance, where features are resistances to different drugs observed in pathogens (Posada-Céspedes et al., 2021; Renz et al., 2025); and other questions in evolutionary biology and disease. EvAMs have a diverse set of forms, from deterministic graph structures describing dependencies between features (acquiring X requires the prior acquisition of Y), and partially ordered sets (posets) describing ordering of feature acquisitions, to transition matrices describing the probability and/or rate of stochastic transitions between different states. All these model forms can fundamentally be pictured as evolution through a state space consisting of all $2^L$ possible states – often with a hypercubic structure linking states (Johnston & Diaz-Uriarte, 2025; Schill et al., 2020).

Different EvAM methods may consider the accumulation process in continuous or discrete time (Aga et al., 2024; Greenbury et al., 2020; Schill et al., 2020); with irreversible or reversible gains of features (Johnston & Diaz-Uriarte, 2025); with dependencies limited to pairwise interactions or relaxed to broader set-wise influences (Aga et al., 2024; Moen & Johnston, 2023); and with independent or correlated (for example, phylogenetically related) samples (as in discrete character evolution (Boyko, 2026; Revell & Harmon, 2022). Across this diverse set, the impact of different covariates on the dynamics of EvAM processes is of great scientific interest – for example, does cancer development follow different pathways in different tumour regions (Khakabimamaghani et al., 2019) or patients, or does drug resistance evolve differently in different lineages (Posada-Céspedes et al., 2021) or geographical regions (Aga et al., 2025)?

Existing methods have addressed the question of how to compare different EvAM inferences. These methods have included comparing predicted trajectories (Diaz-Uriarte & Vasallo, 2019; Luo et al., 2023), conditional predictions of subsequent behaviour (Diaz-Colunga & Diaz-Uriarte, 2021), considerations of identifiability and model equivalence (Diaz-Uriarte, 2018), likelihood- and mixture-model comparisons of conjunctive Bayesian network structures (Gerstung et al., 2009; Montazeri et al., 2016) and related model fitting and parameter comparison (Angaroni et al., 2022; Johnston et al., 2019; Khakabimamaghani et al., 2019). For comparing orderings of feature acquisitions, simple statistical methods like Spearman's rho or Spearman's footrule, Kendall's tau, longest common subsequence, and edit-distance statistics can be used (Alvo & Yu, 2014). For methods inferring posets describing accumulation ordering, the Jaccard index (comparing set intersections and unions) has been applied to compare different model outputs (Posada-Céspedes et al., 2021), including a quantitatively characterised null hypothesis and significance testing capacity. A similar comparison of which features precede which others (sometimes reflected in a "probabilistic feature graph", or a relative ordering matrix) has been used to compare EvAM outputs in other methods (Dauda et al., 2025; Greenbury et al., 2020; Moen & Johnston, 2023). Summaries of transition graphs via an edge filtration provide a multi-layered comparison of different EvAM dynamics (García Pascual et al., 2024). Dimensionality reduction has been used to explore variability in the inferred parameter space in EvAM models and connections with different covariates (Aga et al., 2025; García Pascual et al., 2024; Williams et al., 2013). Notably, the EvAM-Tools platform (Diaz-Uriarte & Herrera-Nieto, 2022) provides a unifying environment for the comparison of different EvAM approaches, encompassing inferred parameters and predicted trajectories.

One issue with some previous methods has been "frameshift" differences in dynamics. Picture two different pathways: 0000-0001-0011-0111-1111 and 0000-1000-1001-1011-1111. There is no overlap in the set of *state* transitions in these two pathways, so transition-based methods like WFCC (García Pascual et al., 2024) will detect a dramatic difference. The feature acquired at each *absolute* ordering is also everywhere different (1,2,3,4 vs 4,1,2,3), so comparisons based on absolute orderings will also report substantial differences (Williams et al., 2013). However, the *relative* orderings of features share substantial similarities (1 before 2, 2 before 3), which will be detected by approaches considering relative orderings (Greenbury et al., 2020; Moen & Johnston, 2023; Posada-Céspedes et al., 2021).

The goal of this note is not to replace or compete with established methods for specific circumstances, but to describe a method supporting the general quantitative comparison of EvAM dynamics that may include the above complications of "frameshift", stochasticity, reversibility, codependence, and multi-step acquisitions. Methods based on deterministic relationships between features cannot be generalised to describe stochastic dependencies. Reversibility immediately challenges several of the above methods, as the possibility that a feature can be lost is not naturally captured in a poset-based or DAG picture, nor an acquisition sequence of fixed length (Johnston & Diaz-Uriarte, 2025). If sets, rather than individual, features can impact the acquisition of other features, binary ordering relationships will not capture dynamics fully (Aga et al., 2024; Moen & Johnston, 2023).

To identify a method that can be applied across these circumstances, we focus on probabilistic summaries of the dynamics supported by a given model, reporting the probability of different prior conditions when a new feature is acquired (Johnston, 2026). Naturally accounting for stochasticity, such an approach can support reversible dynamics and arbitrary influences of sets of features on others, and has a natural parameterisation in terms both of the magnitude of differences in stochastic dynamics and statistical differences between cases. We hope to show that this approach can report robust and interpretable differences in dynamics between different EvAM instances.

## Methods

### *Summarising accumulation dynamics with ordering matrices*

We will use two different methods to summarise inferred dynamics from an EvAM model. The first is the *relative* ordering matrix. Here, $P_{ij}^{(r)}$ gives the probability that feature $i$'s acquisition is preceded by specific feature $j$ across some set of observations (one way of reporting this is considered, for example, in Figs. S2 and S5 of (Dauda et al., 2025)). The second is the *absolute* ordering matrix. Here, $P_{in}^{(a)}$ gives the probability that feature $i$ is acquired after a number $n$ of features are already present (analogous to the cumulative distribution function over "bubble plots" for monotonic EvAM (Aga et al., 2024)).

In each case, an inferred transition matrix is derived. In approaches supporting it, the probability of each transition between states is written down exactly. In sampling-based approaches, the probability of a random walker undergoing a given transition is recorded throughout the sampling process.

Let $I(.)$ be an indicator function returning 1 if its argument is true and 0 otherwise. In the case of irreversible dynamics, a rational choice for a relative ordering matrix is

$$P_{ij}^{(r)} = \sum_t P(t)\, I(s_i^1{=}0)\, I(s_i^2{=}1)\, I(s_j^1{=}1)\,, \qquad \text{(Eqn. 1)}$$

where $P(t)$ is the observation probability of the transition $t$. This expression then accumulates probability for every observed transition involving the acquisition of feature $i$ when feature $j$ is already present. In general, $P_{ij}^{(r)}$ does not necessarily sum to 1 over $i$ or $j$ (for example, if the two traits are acquired simultaneously, or if one is never observed).

For absolute orderings we use:

$$P_{in}^{(a)} = \sum_t P(t)\ I(s_i^1{=}0)\ I(s_i^2{=}1)\ I(\Sigma_k s_k^1 \leq n\text{-}1) \qquad \text{(Eqn. 2)}$$

So that $P_{in}^{(a)}$ gets contributions from every transition where feature $i$ is acquired from a precursor state with at most $n$-1 features present. (The -1 is to allow the first column of the matrix to correspond to the case of zero features present; an additional column for $L$ features present is not needed because no further acquisitions could then occur).

These two matrices are well-defined for cases where accumulation is monotonic and stepwise (one feature acquired at a time), cases where accumulation is reversible, and cases where several features can change simultaneously. In this latter case, $P_{in}^{(a)}$ only considers the number of features in the initial precursor state and does not count contributions from any intermediate states between the precursor and the final state. For example, the transition 001-111 would only contribute probability to features 1 and 2 being acquired after one other feature.

The two matrices serve as summary statistics of inferred dynamics. They can broadly be pictured as describing the probabilities that feature $i$ is acquired after specific feature $j$ ($P_{ij}^{(r)}$) or after at most $n$-1 other features ($P_{\text{in}}^{(a)}$).

### *Statistical comparison of ordering matrices*

Different methods exist in EvAM for quantifying uncertainty in inferred accumulation dynamics. These include Bayesian approaches, where posterior distributions over the governing parameters are sampled (Aga et al., 2024; Greenbury et al., 2020; Johnston & Williams, 2016; Williams et al., 2013) and resampling methods where, for example, the bootstrap is used to generate distributions over parameter estimates (Moen & Johnston, 2023). We will now assume that we have a set of matrices $P_{ij}^{(r,k)}$ and $P_{in}^{(a,k)}$, each corresponding to the $k$th sample from such a set (for example, independent samples from the posterior, or bootstrap resamples). Finally, we will consider two such sets of matrices $P_{ij}^{(r,k,M)}$ and $P_{in}^{(a,k,M)}$, where $M$ now labels different EvAM instances (different models, or different datasets).

If we can demonstrate a systematic difference between the distributions of $P_{ij}^{(r)}$ or $P_{in}^{(a)}$ over the $k$ samples for instance 1 and the $k$ samples for instance 2, we can report a statistical difference between the estimates in the two instances. However, this is necessary but not sufficient for a *scientifically* interesting difference between the two instances. Ordering matrix estimates whose distributions do not overlap, but which do not differ substantially in magnitude, can reflect dynamics which themselves are highly similar. For example, if $P_{12}^{(r,.,1)} \sim 0.2 \pm 0.01$ for instance 1 and $P_{12}^{(r,.,2)} \sim 0.25 \pm 0.01$, there is a lack of overlap between estimates of $P_{12}^{(r)}$ for the two instances, but both instances have a similar probability of feature 1 being acquired before feature 2.

For a statistically robust and scientifically interesting difference, we propose requiring that dynamic estimates are both non-overlapping and reflect different dynamic behaviours. Reported differences will then correspond to *precisely characterised statements about different behaviours*. We propose the following criterion:

$$P(P_{ij}^{(.,.,1)} < q) > 1\text{-}p \text{ and } P(P_{ij}^{(.,.,2)} > 1\text{-}q) > 1\text{-}p \qquad \text{(Eqn. 3)}$$

So that instance 1 has a low probability (<$q$) that $i$ is acquired before $j$, and instance 2 has a high probability (>1-$q$) that $i$ is acquired before $j$, and this condition holds across a proportion 1-$p$ of samples for each instance.

For example, consider $p$=0.2 and $q$=0.25. Enforcing Eqn. 3 will require that >80% of samples of instance 1 have a <25% chance of $i$ being acquired before $j$, and >80% of samples of instance 2 have a >75% chance of $i$ being acquired before $j$. The overlap between the sampled distributions that do not respect the 25/75 separation is then at most $0.2 \times 0.2$=0.04. We will use this $p$=0.2 and $q$=0.25 parameterisation as the default in our examples.

### *Code availability*

This method is included in the *hyperinf* package for EvAM in R (R Core Team, 2022), available at https://github.com/StochasticBiology/hyperinf. The corresponding functions are *ordering_matrix*, *compare_orderings*, and *plot_hyperinf_compare_orderings*. This report additionally makes use of *ape* (Paradis & Schliep, 2019) for tree manipulation and *ggplot2* (Wickham, 2011) and *ggarrange* (Kassambara, 2020) for visualisation. The examples in the manuscript are run from code at https://github.com/StochasticBiology/Compare-EvAM.

## Results

### *Synthetic data*

We first illustrate the form of the ordering matrices in this method for synthetic data. A single evolutionary pathway (first feature 1 is acquired, then feature 2, then 3, then 4) is simulated on a random phylogenetic tree with 32 tips generated from a birth-death process (Fig. 1A) (Giannakis et al., 2024). An irreversible EvAM model (HyperHMM (Moen & Johnston, 2023)) and a reversible one (HyperMk (Johnston & Diaz-Uriarte, 2025)) are used to fit these data. The HyperHMM output is computed over 10 bootstrap resamples of the inferred transitions. The HyperMk model is first fitted to the data, then this fit is used to generate 10 resamples of the fitted model which themselves are fitted, to give a parameteric bootstrap distribution over parameters. The corresponding sets of transition networks are visualised in Fig. 1B, and the two ordering matrices shown for each resampling in Figs. 1C-D. Further comparison of these model outputs is shown in Supp. Fig. 1. Both ordering matrices behave intuitively: $P_{in}^{(a)}$ reports, for example, that feature 1 is overwhelmingly acquired with zero preceding features while feature 4 requires 3 preceding features; $P_{ij}^{(r)}$ reports that feature 1 precedes all others and feature 4 precedes none (with some noise from the resampling of the Mk model).

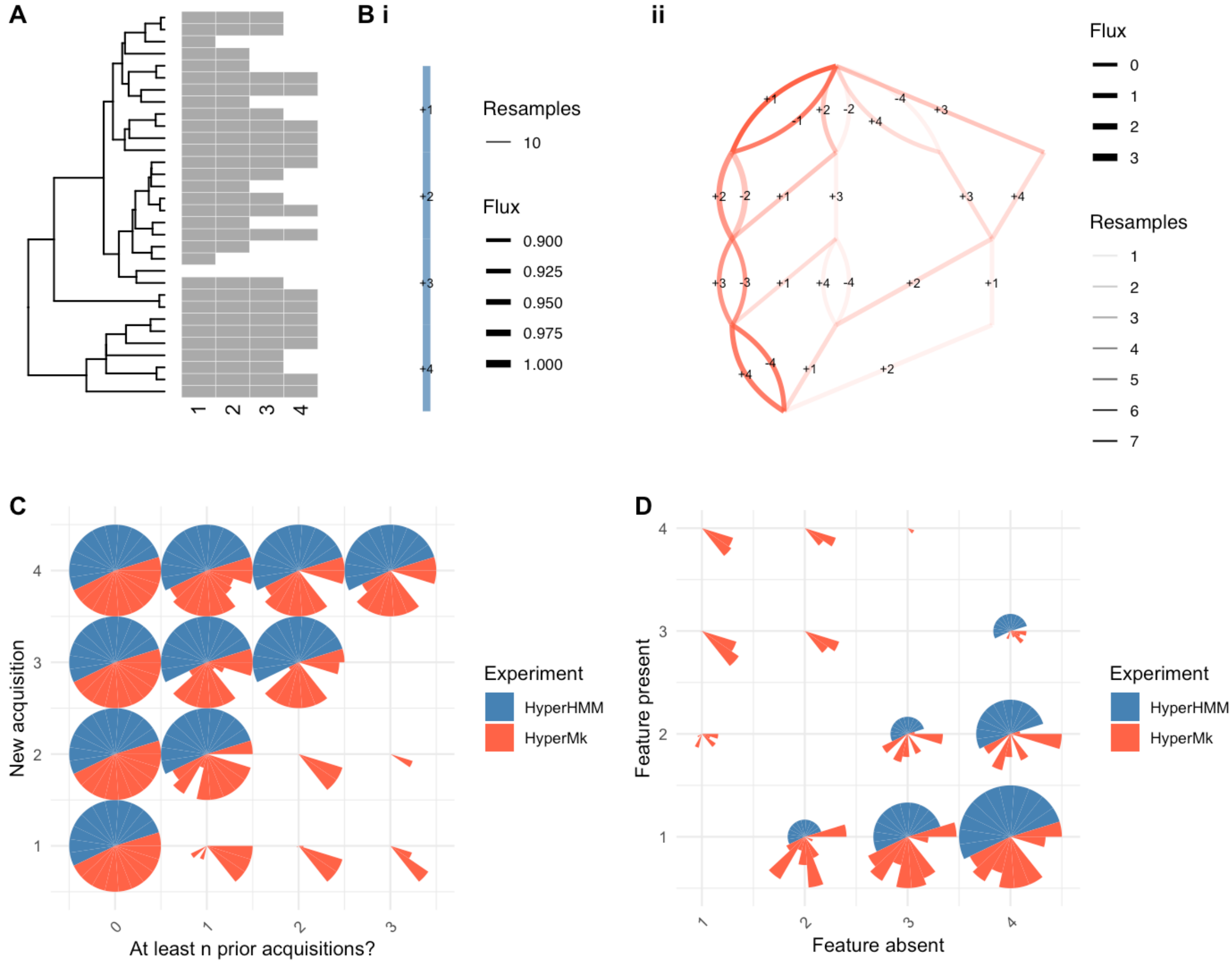

Figure 1. **Ordering matrices for a simple EvAM example. (A)** A synthetic dataset simulated according to a linear pathway (Giannakis et al., 2024): each row is a sample, each column is a feature, with dark pixels corresponding to presence and empty pixels corresponding to absence. The randomly generated phylogeny on the left connects observations (although this approach also works for independent, cross-sectional data). **(B)** Inferred transition networks from (irreversible) HyperHMM (i) and (reversible) HyperMk (ii). Accumulation starts from the top of the figure and proceeds along edges, labelled by the change involved (+ gain; - loss). The opacity of each edge shows how many times it appears among the resampled fits to the data in (A); the width of an edge shows the flux of simulated processes through that edge. **(C)** Absolute ordering matrix $P_{in}^{(a)}$. Each circle segment's radius gives the probability that at least $n$ features are present prior to the acquisition of a given feature. **(D)** Relative ordering matrix $P_{ij}^{(r)}$. Each circle segment's radius gives the probability that feature $j$ is present when feature $i$ is acquired. In both (C) and (D), each segment corresponds to one of 10 resampled fits to the data in (A).

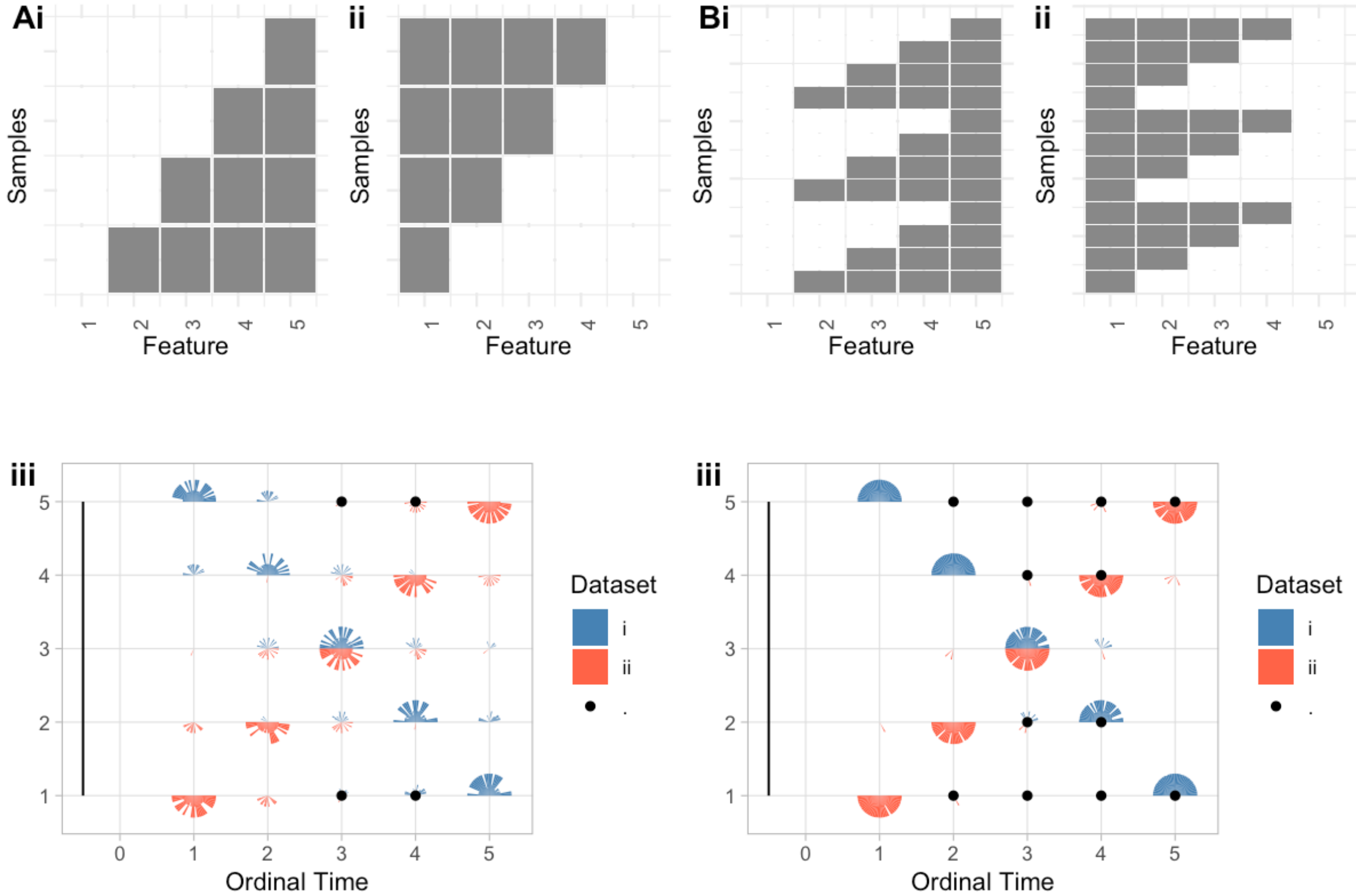

Figure 2. **Detecting differences between inferred EvAM pathways for synthetic data. (A)** Small-data version; **(B)** larger (triplicated) data version. **(i)** and **(ii)** are visualisations of the (cross-sectional) synthetic datasets as in Fig. 1A. **(iii)** summarises the inference and comparison. The radius of each circle segment gives the inferred probability that the given feature (row) is acquired at that timing step (column). Different segments of the same colour correspond to bootstrap resamples; different colours correspond to the different datasets. For example, in Aiii, most bootstrap resamples for dataset 1 give feature 5 a high probability of acquisition in the first step. Vertical lines link features that are identified as having different relative orderings ($P_{ij}^{(r)}$) in the two datasets. For example, feature 1 is robustly inferred to be acquired before feature 5 in dataset 2, and after feature 5 in dataset 1. Black points mark regions where absolute orderings ($P_{in}^{(a)}$) are inferred to be different for a given feature. For example, feature 1 is inferred to have >75% acquisition probability before steps 3 and 4 in dataset 2, while it is inferred to have >75 acquisition probability at or after steps 3 and 4 in dataset 1.

To illustrate the use of this method in comparing EvAM inferences, we next use synthetic data constructed to reflect genuinely different evolutionary pathways. Here, two different datasets (Fig. 2Ai, ii) reflect evolutionary accumulation in distinct pathways: (i) first feature 5, then 4, then 3, and so on; (ii) first feature 1, then 2, then 3, then so on. In Fig. 2A we take a small sample ($n$=4) from these two processes and use HyperHMM with bootstrap resampling (Moen & Johnston, 2023) to infer estimates and uncertainty on these pathway structures. The results are illustrated in Fig. 2Aiii: the distinct pathways are inferred, but uncertainty is considerable. Differences in relative ordering ($P_{ij}^{(r)}$) and differences in absolute ordering ($P_{in}^{(a)}$) are detected for features 1 and 5.

In Fig. 2B we expand the dataset to $n$=12 by triplicating the previous observations. The estimated pathway structure remains consistent, but uncertainty in the inference is dramatically decreased (Fig. 2Biii). Now differences in absolute ordering ($P_{in}^{(a)}$) are also detected for features 2 and 4.

### *Evolution of drug resistance in* Klebsiella pneumoniae *across countries*

To demonstrate the approach on real-world data, we consider the domain of anti-microbial resistance (AMR) evolution, in the particular case of the bacterium *Klebsiella pneumoniae* (Kp) – which we will refer to as KpAMR. KpAMR is a leading global health threat (Naghavi et al., 2024) and the focus of much large-scale genomic research worldwide (Holt et al., 2015). Previous research has used genome data from PathogenWatch and the Kleborate platform (Argimón et al., 2021; Lam et al., 2021) to build a picture of KpAMR evolution across different countries (Aga et al., 2025). Specifically, the EvAM method HyperTraPS (Aga et al., 2024) was used to obtain posterior distributions on the accumulation dynamics of 22 KpAMR features (described in Supp. Table 1) across countries. Specifically, the posterior probability of a feature being acquired at a given step in an ordered evolutionary path (related to our absolute ordering matrix $P_{in}^{(a)}$) was recorded. Principal components analysis of these ordering distributions was used to identify axes of variability across countries, with differences identified, for example, between sub-Saharan African countries and other global regions. The particular instance used to illustrate these differences was between inferred KpAMR dynamics in Gambia and South Korea.

In Fig. 3A, we use our method to test for statistically robust differences in inferred KpAMR dynamics in these two countries. We identify detectable differences in absolute ordering in the features highlighted in the original paper (*Bla_Carb_acquired*, *Flq_mutations*, *Omp_mutations*, *Tet_acquired*), additionally detecting a difference in *Sul_acquired* and *Tmt_acquired* features, demonstrating the power of this approach over less quantitatively targetted methods.

### *Accumulation of chromosomal aberrations in diverse tumour evolution*

EvAM approaches are often applied to infer evolutionary pathways in cancer progression. We obtained chromosomal aberration profiles in a large survey of 2658 cancer across 38 types (Gerstung et al., 2020). For each pair of cancer types, we identified the 8 aberrations most represented in the union of the two profile sets. We ran HyperHMM to infer evolutionary pathways through this space and recorded pairs where our approach reported differences (at the $p$=0.2 and $q$=0.25 level, see Methods) in the corresponding inferred dynamics (for example, Fig. 3B). Fig. 3C shows the set of detected differences and their magnitudes. Several detectable differences occurred between tumour types in the same tissue, or between tissues for the same tumour type. Some examples include CNS-PiloAstro (5p, 5q, 6q acquired earlier) vs CNS-GBM; CNS-Oligo (7q and 7p acquired earlier) vs CNS-PiloAstro later; Kidney-RCC.clearcell (17q and 17p acquired earlier) vs Kidney-RCC.papillary; and Thy-AdenoCA (5q acquired earlier) vs Breast-AdenoCA. Adenocarcinomas were parrticularly variable, with differences observed between many different tissue pairs (Supp. Fig. 2).

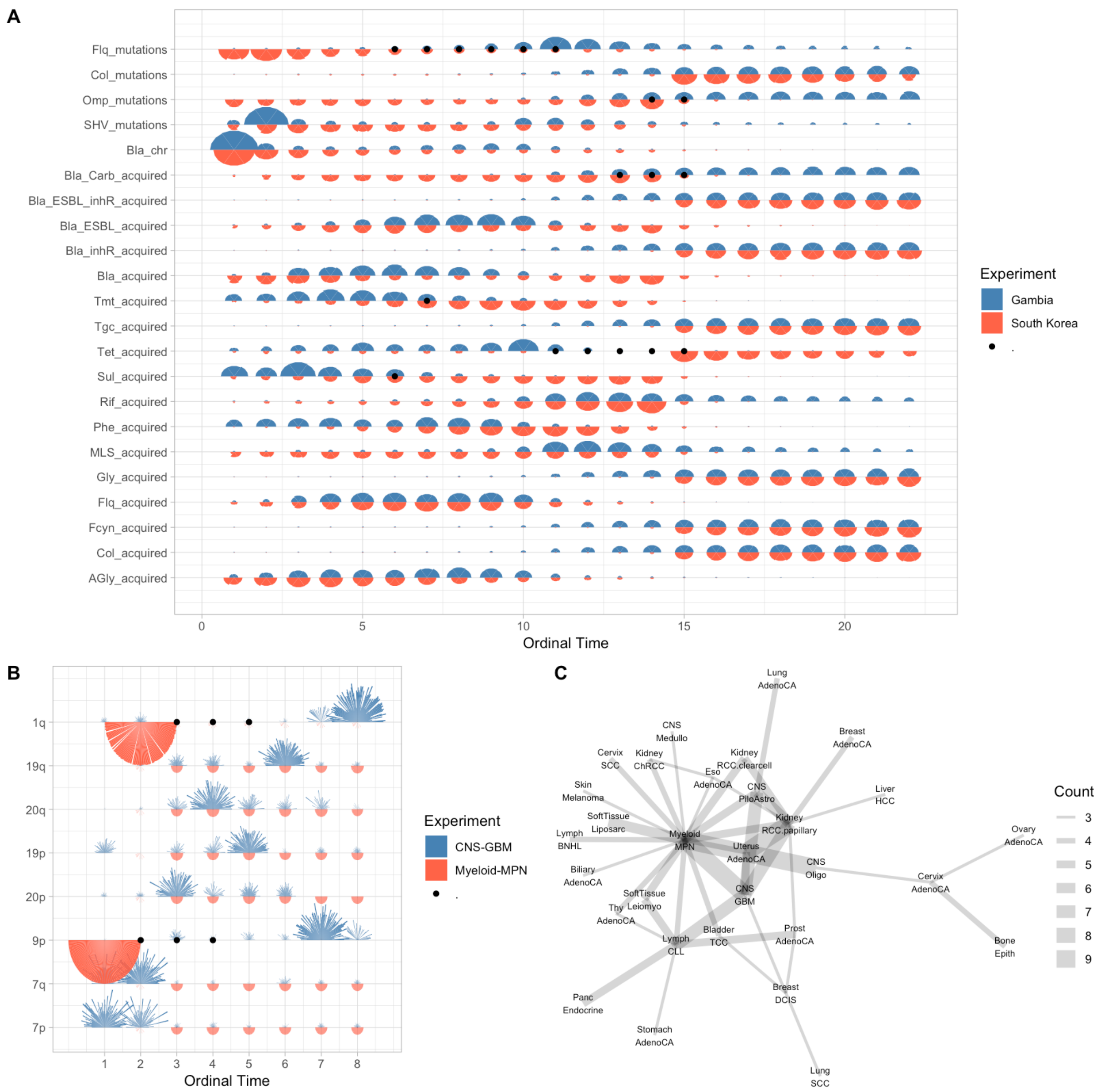

Figure 3. **Real-world examples of differences in inferred EvAM dynamics. (A)** Differences in KpAMR evolution between Gambia and South Korea. See Supp. Table 1 for feature explanations (Aga et al., 2025). As in Fig. 2, black dots define regions where a feature is detected to have robustly different absolute ordering probabilities ($P_{\text{in}}^{(a)}$) – one experiment with >75% probability before the region, the other with >75% probability at or after the region. **(B)** Differences in acquisition of chromosomal aberrations in central nervous system vs myeloid tumours. Features refer to chromosomal regions (chromosome number and p/q arm) where an aberration is acquired. **(C)** Graph of detected differences in accumulation dynamics between different tumour types (abbreviations in Supp. Table 2). An edge between two tumour types denotes a difference detected at the $p$=0.2 and $q$=0.25 level (see Methods); the width of the edge gives the number of such detected differences in the absolute ordering plot (for example, 6 in (B)). For clarity, only tumour pairs with differences detected at more than two positions are shown; the full plot is shown in Supp. Fig. 2.

## Discussion

Our method takes the same basic form as several proposed methods for specific EvAM circumstances – detecting differences in the inferred accumulation dynamics between different cases. Our new contribution here is the flexibility of this particular approach to this goal. In considering probabilistic summaries of the conditions under which a feature is acquired, we support stochasticity, reversibility, higher-order interactions, and potentially multi-step acquisitions in the underlying dynamic model. At the same time we remain conceptually aligned with methods considering, for example, differences in poset summaries of dynamics (Posada-Céspedes et al., 2021), filtrations of fluxes in transition networks (García Pascual et al., 2024), and unsupervised comparison of dynamic summary statistics (Aga et al., 2025; Williams et al., 2013).

This problem is connected to the statistics of rankings and orderings and the comparison of sequences (Alvo & Yu, 2014). The intersection between this stochastic modelling picture and that statistical field will benefit from further exploration. For example, the Plackett-Luce model describes the ordering statistics of independent, exponentially-distributed events, but recent developments have considered a "contextual Plackett-Luce model" where EvAM-like interactions are allowed between events (Mizrachi et al., 2026).

An important distinction that this method highlights (not necessarily for the first time (Diaz-Uriarte, 2018; Diaz-Uriarte & Vasallo, 2019)) is between (i) statistical differences in the parameter inferences from an EvAM model and (ii) notable differences in the dynamics of the EvAM process itself. With large datasets, it is very possible that statistically significant differences exist between estimator distributions for particular parameters in two different models (i). But these may lead only to negligible differences in the dynamics of processes that these models support (ii). The parameterisation of our approach uses bootstrap percentiles to enforce (i) and a given threshold to enforce (ii), allowing tunable control over the importance of the two difference types.

For simplicity, we have focussed on particular single ordering points with different probabilities of acquisition before or after that point. In some cases involving multimodal ordering distributions, the more appropriate focus might be an ordering window (for example, "feature 1 has >75% probability of being acquired before step 3 or after step 7; feature 2 has a <25% probability of being acquired in this range"). The idea of comparing resampled probabilities can readily be generalised to this situation, with a corresponding increase in the search space of possible windows of discrepancy.

In addition to the case studies on KpAMR and cancer evolution above, previous work has used different EvAM comparison approaches to explore differences in the evolution of: HIV drug resistance in different lineages (Posada-Céspedes et al., 2021); tool use in different animal lineages and environments (Johnston & Røyrvik, 2020), photosynthesis in different plant lineages (Williams et al., 2013). The collection of applications to cancer progression is broad enough to have motivated several review articles (Beerenwinkel et al., 2015; Diaz-Uriarte & Johnston, 2025; Diaz-Uriarte & Vasallo, 2019). Applying EvAM to disease progression, (Johnston et al., 2019) explored differences in the accumulation of clinical symptoms in malaria for different prognoses (survival vs death); a predictive model based on these differences was generated and validated for new observations. We hope that the flexible approach reported here will help formalise EvAM comparisons across these diverse domains in future applications.

## Acknowledgements

This work was supported by the Trond Mohn Foundation (project HyperEvol under grant agreement No. TMS2021TMT09 to I.G.J.) through the Centre for Antimicrobial Resistance in Western Norway (CAMRIA) (TMS2020TMT11). This project has received funding from the European Research Council (ERC) under the European Union's Horizon 2020 research and innovation program (grant agreement No. 805046 [EvoConBiO] to I.G.J). This work was supported by grant PID2024-156888OB-I00 funded by MICIU /AEI/10.13039/501100011033 / FEDER, EU to R.D-U.

## Supplementary Information

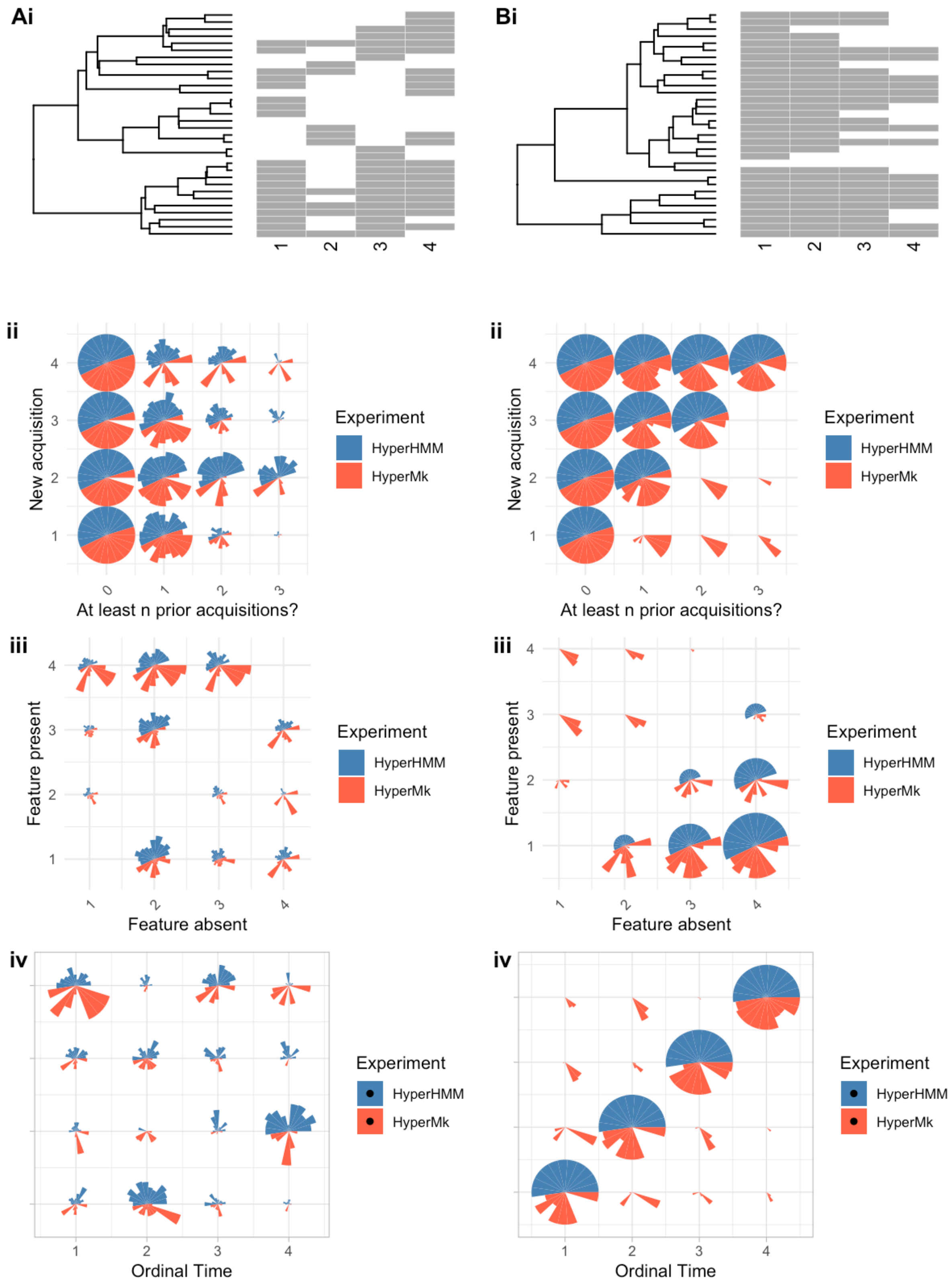

Supplementary Figure 1. **Comparison of (irreversible) HyperHMM and (reversible) HyperMk approaches on different synthetic datasets.** (A) Data simulated for random, independent features; (B) data simulated from the single pathway model in Fig. 1, each on a random birth-death tree with 32 tips. (i) Visualisation of the datasets; (ii) absolute ordering matrices; (iii) relative ordering matrices; (iv) comparison plots for inference with HyperHMM (10 bootstrap resamples of transitions); HyperMk (10 parameteric bootstrap resamples from resimulated data from the fitted model).

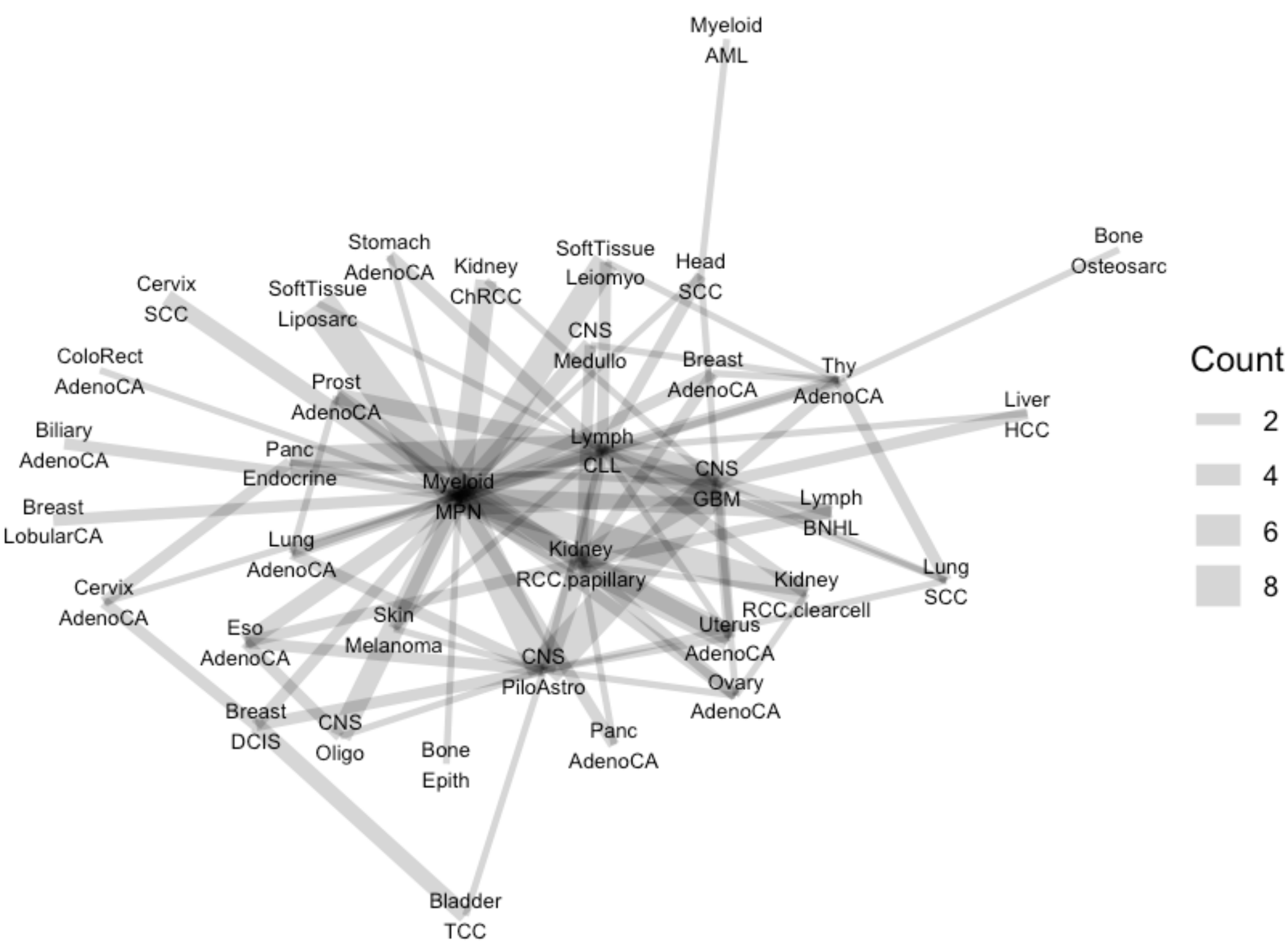

Supplementary Figure 2. **Differences in cancer EvAM.** Full network of detected differences in accumulation dynamics between cancer types (including Myeloid MPN). As in Fig. 3C, an edge between two tumour types signals a robustly detected different in accumulation dynamics; the width of the edge gives the number of detected differences in the absolute ordering plot. Abbreviations in Supp. Table 2.

Supplementary Table 1. **KpAMR character names.** More details available via Kleborate documentation at https://github.com/klebgenomics/Kleborate and (Aga et al., 2025).

| **Abbreviation** | **Resistance to drug type (or other feature)** |
|---|---|
| *AGly* | Aminoglycosides |
| *Col* | Colistin |
| *Fcyn* | Fosfomycin |
| *Flq* | Fluoroquinolones |
| *Gly* | Glycopeptides |
| *MLS* | Macrolides |
| *Phe* | Phenicols |
| *Rif* | Rifampin |
| *Sul* | Sulfonamides |
| *Tet* | Tetracyclines |
| *Tgc* | Tigecycline |
| *Tmt* | Trimethoprim |
| *Bla_a* | β-lactam resistance without extended spectrum or inhibitor-resistance |
| *Bla_inhR* | Resistance to β-lactam/inhibitor combinations |
| *Bla_ESBL* | Resistance to extended-spectrum β-lactams |
| *Bla_ESBL_inhR* | Resistance to extended-spectrum β-lactam/inhibitor combinations |
| *Bla_Carb* | Resistance to carbapenems |
| *SHV* | SHV β-lactamase with expanded enzyme activity |
| *Bla_chr* | SHV alleles conferring resistance to ampicillin |
| *Omp* | Outer membrane protein |

Supplementary Table 2. **Tumour character names.** More details available via (Gerstung et al., 2020).

| **Abbreviation** | **Full form** |
|---|---|
| *AdenoCA* | Adenocarcinoma |
| *HCC* | Hepatocellular carcinoma |
| *BNHL* | B-cell non-Hodgkin lymphoma |
| *RCC.clearcell* | Renal cell carcinoma (clear cell type) |
| *SCC* | Squamous cell carcinoma |
| *Medullo* | Medulloblastoma |
| *GBM* | Glioblastoma multiforme |
| *Endocrine* | Neuroendocrine carcinoma |
| *Leiomyo* | Leiomyosarcoma |
| *Liposarc* | Liposarcoma |
| *ChRCC* | Chromophobe renal cell carcinoma |
| *CLL* | Chronic lymphocytic leukemia |
| *TCC* | Transitional cell carcinoma |
| *RCC.papillary* | Papillary renal cell carcinoma |
| *PiloAstro* | Pilocytic astrocytoma |
| *Oligo* | Oligodendroglioma |
| *LobularCA* | Invasive lobular carcinoma |
| *DCIS* | Ductal carcinoma in situ |
| *AML* | Acute myeloid leukemia |
| *MPN* | Myeloproliferative neoplasm |
| *MDS* | Myelodysplastic syndrome |
| *Epith* | Epithelial tumor |